\documentclass[psf,article,submit,pdftex,moreauthors]{Definitions/mdpi} 

\firstpage{1} 
\pubvolume{1}
\issuenum{1}
\articlenumber{0}
\pubyear{2024}
\copyrightyear{2024}
\datereceived{ } 
\daterevised{ } % Comment out if no revised date
\dateaccepted{ } 
\datepublished{ } 
\hreflink{https://doi.org/} % If needed use \linebreak
\Title{Nested Sampling for Exploring Lennard-Jones Clusters}

\TitleCitation{Nested Sampling for Exploring Lennard-Jones Clusters}

\Author{Lune Maillard $^{1,}$*, Fabio Finocchi $^{1}$, César Godinho$^{2}$ and Martino Trassinelli $^{1,}$*}

\AuthorNames{Lune Maillard, Fabio Finocchi, César Godinho and Martino Trassinelli}

\AuthorCitation{Maillard, L.; Finocchi, F.; C. Godinho ; Trassinelli, M.}
\address{%
$^{1}$ \quad Sorbonne Université, CNRS, Institut des Nanosciences de Paris, INSP, F-75005 Paris, France\\ %Institut des Nanosciences de Paris, Sorbonne Université, CNRS, France
$^{2}$ \quad Laboratory of Instrumentation, Biomedical Engineering and Radiation Physics (LIBPhys-UNL), Department of
Physics, NOVA School of Science and Technology, NOVA University Lisbon, Caparica, Portugal\\}

\corres{Correspondence: lune.maillard@insp.upmc.fr (L.M.), martino.trassinelli@insp.jussieu.fr (M.T)}

\abstract{Lennard-Jones clusters, while an easy system, have a significant number of non equivalent configurations that increases rapidly with the number of atoms in the cluster. Here, we aim at determining the cluster partition function; we use the nested sampling algorithm, which transforms the multidimensional integral into a one-dimensional one, to perform this task. In particular, we use the \texttt{nested\_fit} program, which implements slice sampling as search algorithm. We study here the 7-atom and 36-atom clusters to benchmark \texttt{nested\_fit} for the exploration of potential energy surfaces. We find that \texttt{nested\_fit} is able to recover phase transitions and find different stable configurations of the cluster. Furthermore, the implementation of the slice sampling algorithm has a clear impact on the computational cost.}

\keyword{Lennard-Jones; Clusters; Nested sampling} 

\begin{document}
\nolinenumbers
%%%%%%%%%%%%%%%%%%%%%%%%%%%%%%%%%%%%%%%%%%
\section{Introduction}
Lennard-Jones clusters are relevant models for rare gas atoms clusters \cite{ashcroft_solid_1976} but the number of configurations increases exponentially with the number of degrees of freedom \cite{doye_evolution_1999}. They have been widely studied using a variety of methods such as basin-hopping \cite{wales_global_1997}, Metropolis Monte-Carlo \cite{quirke_melting_1984}, Molecular Dynamics \cite{beck_interplay_1988} or parallel tempering \cite{neirotti_phase_2000,calvo_phase_2000}.

In this work, we use the nested sampling algorithm \cite{skilling_nested_2004} to study those clusters. This algorithm is mainly employed to analyse data in fields such as cosmology and astrophysics \cite{ashton_nested_2022} but is applied more and more in materials science for the exploration of potential energy surfaces, including Lennard-Jones clusters \cite{partay_efficient_2010}. Here, we utilise \texttt{nested\_fit} \cite{trassinelli_nested_fit_2019, trassinelli_bayesian_2017, trassinelli_mean_2020, maillard_assessing_2023}, which has a different implementation of nested sampling than the one used in Ref. \cite{partay_efficient_2010}. Indeed, we want to benchmark \texttt{nested\_fit} for the exploration of potential energy surfaces for classical and, in the future, quantum systems.

We first present the implementation of the nested sampling algorithm for the specific case of Lennard-Jones cluster in Section 2. In Section 3, we study, in particular, clusters of two sizes. In Section 4, we look at the computational cost of our implementation. Finally, we conclude in Section 5. 

\section{Nested sampling for Lennard-Jones clusters}
For a cluster of $N$ atoms, the truncated and shifted Lennard-Jones potential takes the following form \cite{shi_histogram_2001,partay_polytypism_2017}
\begin{equation*}
V=\sum_{\substack{1\leq k < j \leq N\\r_{i,kj}<r_c}} 4\epsilon\left(\left(\frac{r_0}{r_{ij}}\right)^{12}-\left(\frac{r_0}{r_{ij}}\right)^{6}-\left(\frac{r_0}{r_c}\right)^{12}+\left(\frac{r_0}{r_c}\right)^{6}\right).
\end{equation*}
We have that $r_{ij}$ is the distance between atoms $i$ and $j$ and that $r_c$ is the cutoff radius, used to remove the interaction at infinite range. 
The partition function for this system can be written as

\begin{equation*}\label{eq_ev_vpos}
\begin{split}
Z(\beta)&=\frac{1}{N!~h^{3N}}\int \exp\left(-\beta \left(\frac{1}{2}\sum_{i=1}^{3N}\frac{p_i^2}{m}+V(\boldsymbol{x})\right)\right) d\boldsymbol{x} d\boldsymbol{p}\\
&=\frac{1}{N!~h^{3N}}\int \exp\left(-\beta \left(\frac{1}{2}\sum_{i=1}^{3N}\frac{p_i^2}{m}\right)\right) d\boldsymbol{p}\int \exp(-\beta (V(\boldsymbol{x})) d\boldsymbol{x}=Z_k*Z_c,   
\end{split}
\end{equation*}
% \begin{equation*}\label{eq_ev_vpos}
% Z(\beta)=\frac{1}{N!~h^{3N}}\int \exp\left(-\beta \left(\frac{1}{2}\sum_{i=1}^{3N}\frac{p_i^2}{m}+V(\boldsymbol{x})\right)\right) d\boldsymbol{x} d\boldsymbol{p}=Z_k*Z_c,   
% \end{equation*}
where $\boldsymbol{p}=(p_1,...,p_{3N})$ is the vector of momenta, $\boldsymbol{x}$ the vector of all atomic positions, $\beta=\frac{1}{k_BT}$ the inverse temperature and $h$ the Planck constant. The separation in the equation above is possible as the potential $V$ only depends on the atom positions and not on the momenta. The term $Z_k=\frac{1}{N!~h^{3N}}\int \exp\left(-\beta \left(\frac{1}{2}\sum_{i=1}^{3N}\frac{p_i^2}{m}\right)\right) d\boldsymbol{p}$ is easy to compute and gives $(\sqrt{2\pi m/(\beta h^2)})^{3N}/N!$. The term $Z_c=\int\exp(-\beta (V(\boldsymbol{x})) d\boldsymbol{x}$ can be rewritten as $Z_c=\int\rho(E)\exp(-\beta E) dE$ where $\rho(E)$ is the density of states \cite{partay_nested_2021} and does not in general have an analytical form. We compute $Z_c$ by using a Monte Carlo sampling and, more precisely, the nested sampling algorithm which works in the following way \cite{skilling_nested_2004,partay_efficient_2010}:
\begin{enumerate}
    \item $K$ points, called live points, are uniformly sampled from the entire space.
    \item At each iteration $i$, the point $\boldsymbol{x}_{old}$ associated to the highest energy is removed and replaced by a point $\boldsymbol{x}_{new}$ with an energy that is strictly lower: $V(\boldsymbol{x}_{new})<V(\boldsymbol{x}_{old})$. The method to find this new point will be presented in Section 2.1. Denoting $E_i=V(\boldsymbol{x}_{old})$, this point will contribute to the partition function as
    \begin{equation}\label{eq_weight}
     c_i=w_ie^{-\beta E_i} \mbox{ with }w_i=\frac{1}{2}\left(\left(\frac{K}{K+1}\right)^{m-1}-\left(\frac{K}{K+1}\right)^{m+1}\right).   
    \end{equation} The term $w_i$ is the approximation of the density of states (DOS) that is evaluated by statistical considerations.
    \item This procedure is repeated until the current contribution is small compared to previous contributions: $\log(c_i)-\log(c_{max})<\delta$ with $c_{max}=\max_{m\leq i}(c_m)$ at an inverse temperature $\beta$ chosen by the user. Here, we use $\delta=-10$, the value used in Ref. \cite{maillard_assessing_2023} that gave good results for the harmonic potential.
\end{enumerate}
As the interaction potential does not depend on the temperature, we can compute the partition function at all temperatures by estimating pairs of ($E_i$,$w_i$) values during one single exploration of the potential \cite{partay_efficient_2010} as
\begin{equation*}\label{eq_ev_calc_form_x}
Z_c(\beta)\approx\sum_i c_i.   
\end{equation*}

From the partition function, other properties of the system can be calculated such as the internal energy 
\begin{equation*}
U=-\frac{\partial \log(Z)}{\partial \beta}=-\frac{\partial \log(Z_k)}{\partial \beta}-\frac{\partial \log(Z_c)}{\partial \beta}=\frac{3N}{2}k_BT-\frac{\partial \log(Z_c)}{\partial \beta}
\end{equation*}
and the heat capacity
\begin{equation*}
C_v=\frac{\partial U}{\partial T}=\frac{3N}{2}k_B-\frac{\partial }{\partial T}\frac{\partial \log(Z_c)}{\partial \beta}.    
\end{equation*}

In this paper, we use, in particular, the \texttt{nested\_fit} program. In the next sections, we present some of the specific implementation of nested sampling in \texttt{nested\_fit}.

\subsection{Slice sampling real and slice sampling transformed}
The method used to find the new point $\boldsymbol{x}_{new}$ in \texttt{nested\_fit} is slice sampling \cite{neal_slice_2003,handley_polychord_2015-1}. This method consists in uniformly choosing new exploration points on a slice of the volume defined by the constraint $V(\boldsymbol{x})<V(\boldsymbol{x}_{old})$. In one dimension, one of the live points is randomly chosen and a slice is built around it until the end points of the slice have an energy higher than $\boldsymbol{x}_{old}$ or are out of the sampling space. A point is then sampled from within the slice and accepted if it verifies the constraint and rejected otherwise. In that case, another point is sampled until an acceptable point is found. In a multidimensional setting, a change of coordinates is first performed to efficiently explore all the parameter space, even in presence of strong correlation. This transformation is done via the Cholesky transformation of the covariance matrix of the live points of the considered step to transform the points coordinates into new coordinates with dimensions $\sim\mathcal{O}(1)$ in all directions \cite{handley_polychord_2015-1}. The one-dimensional algorithm is then applied recursively to the vectors of $n_\textrm{bases}$ randomly generated orthonormal bases (see Ref. \cite{maillard_assessing_2023}). Hence, to compute an update candidate, one live point, corresponding to one $N$-atom cluster, is moved for $n_\textrm{bases}\times 3N$ steps, each one moving all atoms in the cluster. Increasing the value of $n_\textrm{bases}$ implies that there is a better decorrelation of samples but it also results in a higher computational cost.

There are two ways of implementing this algorithm:
\begin{enumerate}
    \item First, the steps can be performed in the transformed space. This will be referred to as \emph{slice sampling transformed} and was used in Ref. \cite{maillard_assessing_2023}. However, the sampled points need to be transformed back to the real space to compute the energy and check that the points are within the bounds. This brings a significant computational overhead to the computation.
    \item Second, the steps can be performed in the real space and only the slices are chosen in the transformed space. This will be referred to as \emph{slice sampling real} and was used in Refs. \cite{handley_polychord_2015,handley_polychord_2015-1}. In this case, only the orthonormal bases need to be transformed to the real space.
\end{enumerate}
\subsection{Parallelisation}
The nested sampling algorithm can be easily parallelised. One method that is used to make nested sampling parallel is, instead of searching and substituting one point per iteration, to search $r$ new points in parallel and then to replace $r$ points at once \cite{henderson_parallelized_2014}: the $r$ points with the highest energy are removed and replaced by $r$ new points with an energy lower than the energies of all removed points. In that case, the compression factor is $\frac{K-r+1}{K}$ instead of $\frac{K}{K+1}$ (see Eq. \eqref{eq_weight}) \cite{partay_nested_2021}. However, this parallelisation increases the variance of the results \cite{partay_nested_2021}. A different approach is used in \texttt{Polychord} using MPI protocols \cite{handley_polychord_2015,handley_polychord_2015-1}: one primary process is in charge of the sequential replacing of the live points while all other secondary processes are in charge of finding new live points continuously, with the threshold criterion changed by the primary process. The sequential replacing is thus done by the primary process with the points that are found by the secondary processes. A point found for an iteration with a specific energy constraint is valid for a further iteration if it verifies the new constraint \cite{handley_polychord_2015-1}. In this case, the compression factor is still $\frac{K}{K+1}$.

The method we use is very similar to the one used in \texttt{Polychord}. First, the point with highest energy $\boldsymbol{x}_{max}$ is identified; we note its associated energy $V_{max}$. Then, $r$ searches are done in parallel to find a new point with energy lower than $V_{max}$. We thus obtain a list of $r$ new points each with its associated energy. Finally, the points are sequentially added (without parallelisation) to the set of live points by comparing their energy with $V_{max}$ which evolves with the addition of the new points. There are two possibilities:
\begin{itemize}
    \item If the energy is lower than $V_{max}$, the point is added to the live points, $\boldsymbol{x}_{max}$ is removed and $V_{max}$ is updated to the new maximum energy of the set of live points. The new value will be used for new points not yet added or rejected.
    \item If the energy is higher than $V_{max}$, the new point is rejected and $V_{max}$ is not updated.
\end{itemize}
Hence, for each set of searches done, between one and $r$ points are replaced in our set of live points. Indeed, by definition, all the searches have an energy lower than the initial $V_{max}$ so that at least one point is added.
\subsection{Computing the covariance matrix}
Another aspect of \texttt{nested\_fit} that brings a computational overhead is the computation of the covariance matrix and its Cholesky transform at every iteration. To accelerate the program, we have chosen to compute those two matrices only every $0.05K$ iterations. This value was chosen as a trade-off between accuracy and computational cost.
\section{Lennard-Jones clusters}

We now study Lennard-Jones clusters of two sizes : 7 atoms and 36 atoms. In the case of the 7-atom cluster, we compare our results with those found in Ref. \cite{partay_efficient_2010}. In this work, we use reduced units i.e. the temperature is in units of $\frac{k_B T}{\epsilon}$ with $\epsilon=1$. In reduced units, the stopping temperature is going to be $0.01$ for the 7-atom cluster and $0.005$ for the 36-atom cluster. For the simulation, the atoms are confined in a cubic box of side $L$. To fix the size of the box, we will look at the value of the density $\rho$ in unit of $r_0^{-3}$: $\rho=\frac{N}{(L/r_0)^3}r_0^{-3}$. In all our cases, we take $L=6$ and use $r_0$ to tune the density to a given desired value. Within reduced units, the results are valid for all atomic species interacting via the Lennard-Jones potential: the specific ($r_0$,$\epsilon$) values, corresponding to the atomic species considered, need to be reintroduced instead of the reduced ones. Hence, the specific value of $r_0$ is not important. The density characterises the size of the space the particles evolve in, relative to the number of atoms $N$. We use slice sampling real with the covariance matrix computed every $0.05K$ iterations and the procedure is repeated eight times to estimate the uncertainties.

Here, we aim to compute the partition function and derivatives of the clusters, aiming at detecting the possible phase transitions. In an infinite system, a phase transition is indicated by a discontinuity in the heat capacity curve, either in the form of a singularity (delta function) or a critical exponent \cite{huller_first_1994}. However, here we work in a finite system composed of $N$ particles in which case a phase transition is not shown by a discontinuity but rather by a peak (or a shoulder) in the heat capacity curve \cite{medved_modeling_2017}. %This peak can sometimes take the form of a shoulder.

\subsection{7 atoms}
\begin{figure}
    \captionsetup{justification=centering}
\begin{adjustwidth}{-\extralength}{0cm}
    \centering
    \includegraphics[scale=0.33,trim=0 10 0 5, clip]{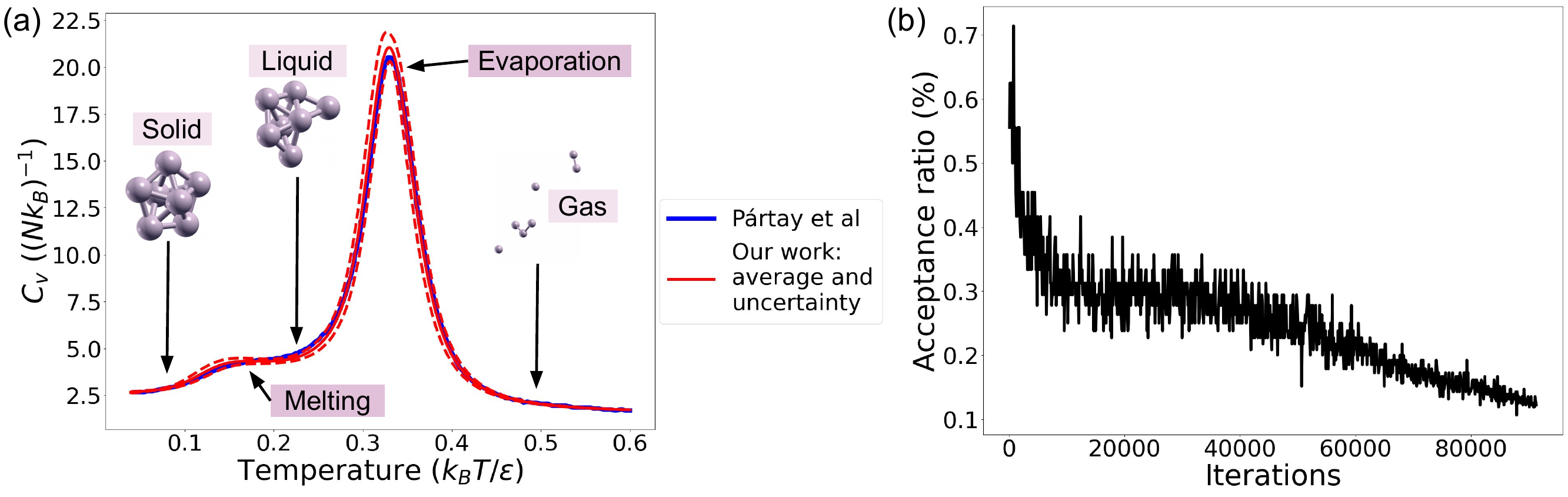}
\end{adjustwidth}
    \caption[Classical Lennard-Jones 7]{(a) Heat capacity for the Lennard-Jones cluster with $7$ atoms. Comparison with the results in Ref. \cite{partay_efficient_2010}. The configurations shown were obtained with \texttt{XCrySDen} \cite{kokalj_xcrysdennew_1999}. (b) Evolution of slice sampling's acceptance ratio during one of the eight runs.}
    \label{lj_7_class}    
\end{figure}
First, we consider a cluster made of $N=7$ atoms with a density of $\rho=9.27\times10^{-3}r_0^{-3}$ (corresponding to $r_0=0.659$). The size of the cluster and the density are chosen in order to compare our results with those obtained in Ref. \cite{partay_efficient_2010}. We use $K=1000$ live points and $n_\textrm{bases}=5$ so that the samples are decorrelated enough for the comparison. Ref. \cite{partay_efficient_2010}, through the \texttt{pymatnest} \cite{partay_efficient_2010,baldock_determining_2016,baldock_constant-pressure_2017,partay_nested_2021} program, moves the atoms one at a time using rejection Gibbs sampling with $K=500$ live points. We both use a cutoff radius $r_c$ of $3r_0$. Both heat capacity curves obtained are represented in Figure \ref{lj_7_class} (a) with examples of solid, liquid and gas configurations. We see that a good agreement with Ref. \cite{partay_efficient_2010} is obtained. More importantly, \texttt{nested\_fit} is able to recover both the evaporation and melting phase transitions. We also ran \texttt{nested\_fit} with $K=500$ live points but obtained an evaporation peak that was slightly shifted towards higher temperatures compared to the curve from Ref. \cite{partay_efficient_2010}, indicating that there are too few points to obtain convergence. The difference in search method may explain why we need to double the points compared to Ref. \cite{partay_efficient_2010}. Indeed, \texttt{pymatnest} was developed specifically for the exploration of energy surfaces while \texttt{nested\_fit} has applications in both data analysis and materials science. The search method used by the latter thus have to work in both cases. Furthermore, in Ref. \cite{partay_efficient_2010}, the Lennard-Jones function was called $2\times10^7$ times while here it was called $5.4\times10^7$ times. There are two main reasons for this increase: first, we use twice as many live points, which should more or less double the number of calls, and second, in slice sampling, the function is called not only to move the points but also to extend the slice. In Figure \ref{lj_7_class} (b), we see that the acceptance ratio of a slice sampling search decreases when the volume decreases, showing that it is then more difficult to find new sampling points. The explorations were done on a 64 bits computer with 4 CPUs of frequency 3.3 GHz and took around one minute per exploration.

\subsection{36 atoms}
\begin{figure}
    \captionsetup{justification=centering}
    \centering
    \includegraphics[scale=0.4,trim=0 5 0 5, clip]{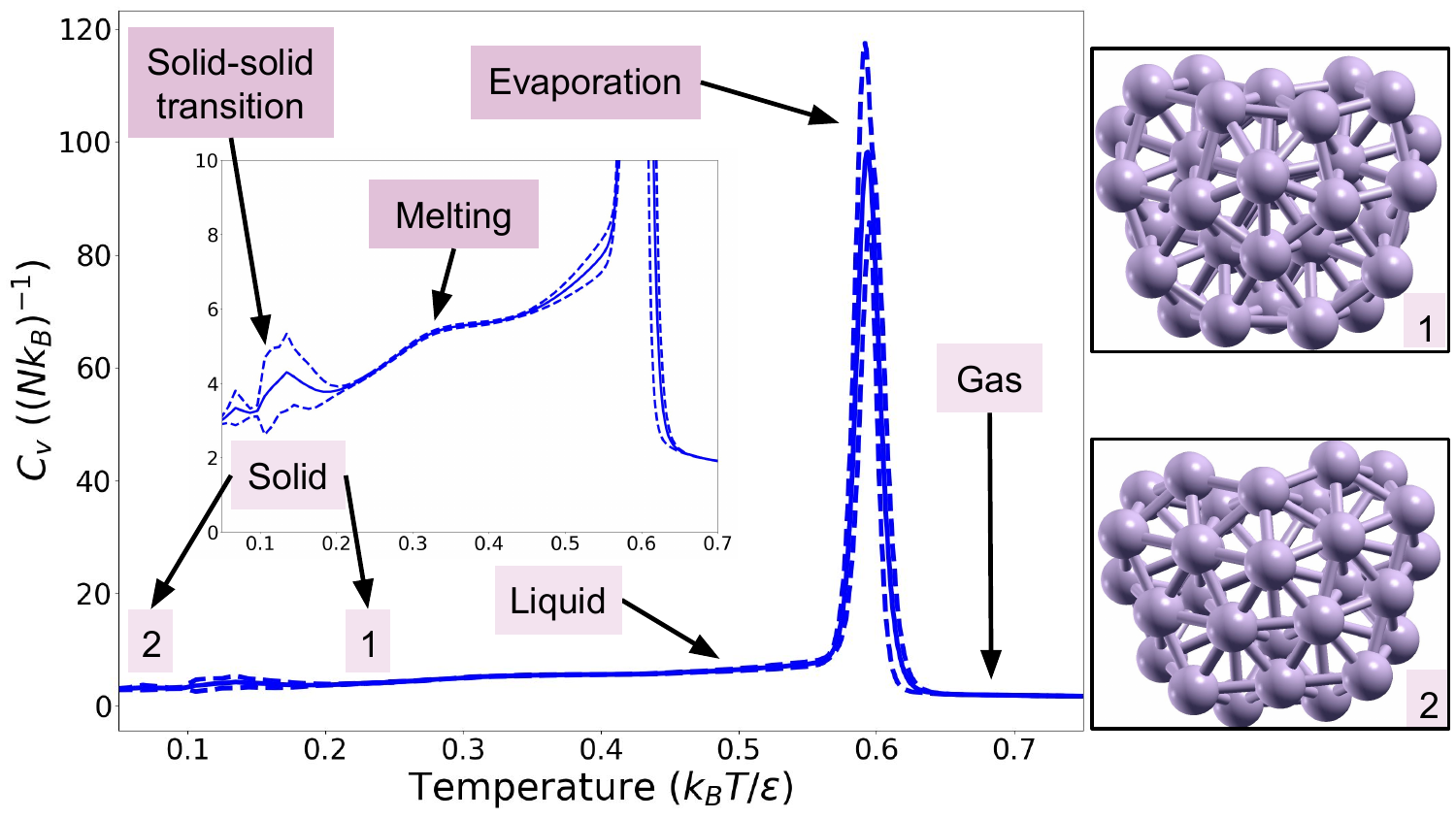}
    \caption[Classical Lennard-Jones 36]{Heat capacity for the Lennard-Jones cluster with $36$ atoms. The configurations shown were obtained with \texttt{XCrySDen} \cite{kokalj_xcrysdennew_1999}. The full line is the mean obtained from the eight runs and the dashed curves correspond to the first standard deviation.}
    \label{lj_36_class}    
\end{figure}
Second, we consider a cluster composed of $N=36$ atoms with a density of $\rho=1.79\times10^{-2}r_0^{-3}$ (corresponding to $r_0=0.475$). This example is quite challenging as a low temperature peak is observed, associated to a Mackay - anti-Mackay phase transition \cite{mandelshtam_multiple_2006,frantsuzov_size-temperature_2005}. Here, we use $n_\textrm{bases}=3$ as the energy function is more costly than in the previous case. The heat capacity obtained with $K=70000$ is shown in Figure \ref{lj_36_class}. Again, we can see the presence of a peak corresponding to evaporation and a shoulder corresponding to melting. However, in this case, there is also a peak at lower temperature (around $T=0.135$) that corresponds to a solid-solid phase transition. From the standard deviation of the heat capacity curve, we can see that the position of this peak is more unstable from one run to the other than the position of the evaporation peak. Indeed, this low temperature peak is harder to obtain --- and was not seen in runs with fewer live points --- as we need enough points to fall in each basin so that it is correctly explored. A variation in the proportion of samples in each basin could explain this difference between the runs. We found the positions of the peaks to be similar when more live points --- up to $K=110000$ --- were used, indicating that $K=70000$ live points is enough to attain convergence. The two solid configurations are shown in Figure \ref{lj_36_class}. The low temperature peak was also found in Refs. \cite{partay_efficient_2010, mandelshtam_multiple_2006, frantsuzov_size-temperature_2005} at around $T=0.145$ which is in the interval $[0.1,0.15]$ given by our eight runs. The study in Refs. \cite{partay_efficient_2010, mandelshtam_multiple_2006, frantsuzov_size-temperature_2005} were done at different densities, but this does not affect the position of the solid-solid transition peak. We also tried using $n_\textrm{bases}=1$ but we observed a shift of the evaporation peak: the number of bases used is probably too small to decorrelate the samples. The explorations were done on a 64 bits computer with 64 CPUs of frequency 2.0 GHz and took around 43 hours per exploration.

\section{Program profiling and optimisation}
\begin{figure}
\begin{adjustwidth}{-\extralength}{0cm}
    \centering
    \includegraphics[scale=0.5,trim=0 232 0 0, clip]{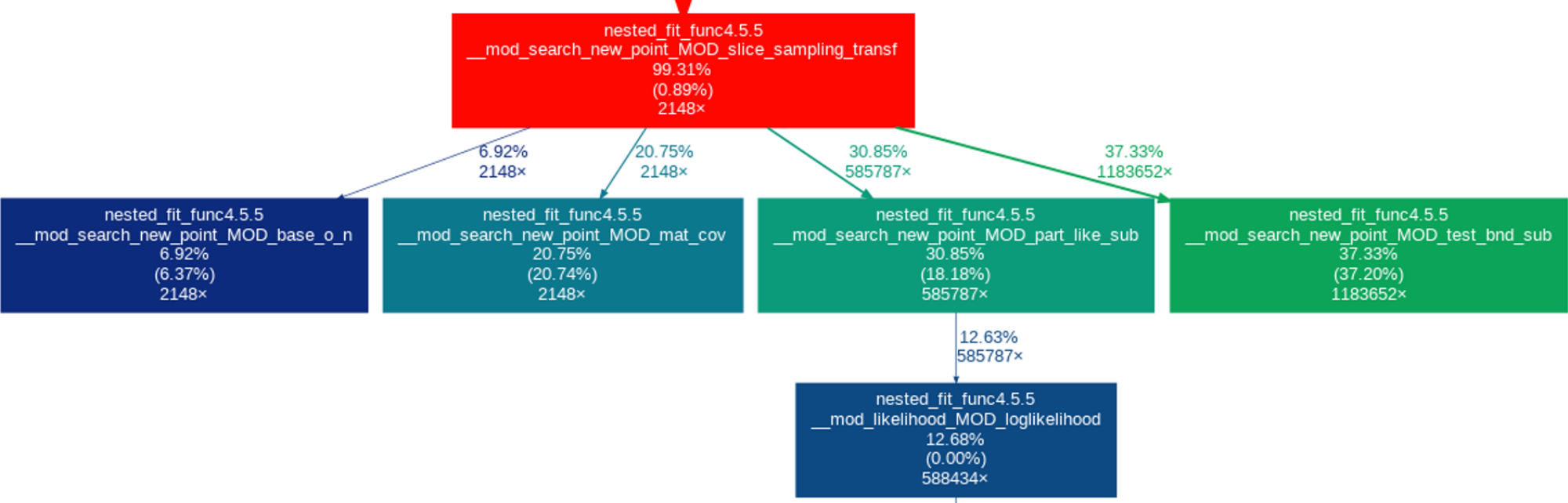}
\end{adjustwidth}
    \caption{Profiling for slice sampling transformed with the calculation of the covariance matrix at each iteration}    
    \label{fig_profiling_sst_full}
\end{figure}
\begin{figure}
\begin{adjustwidth}{-\extralength}{0cm}
    \centering
    \includegraphics[scale=0.4,trim=0 135 0 0, clip]{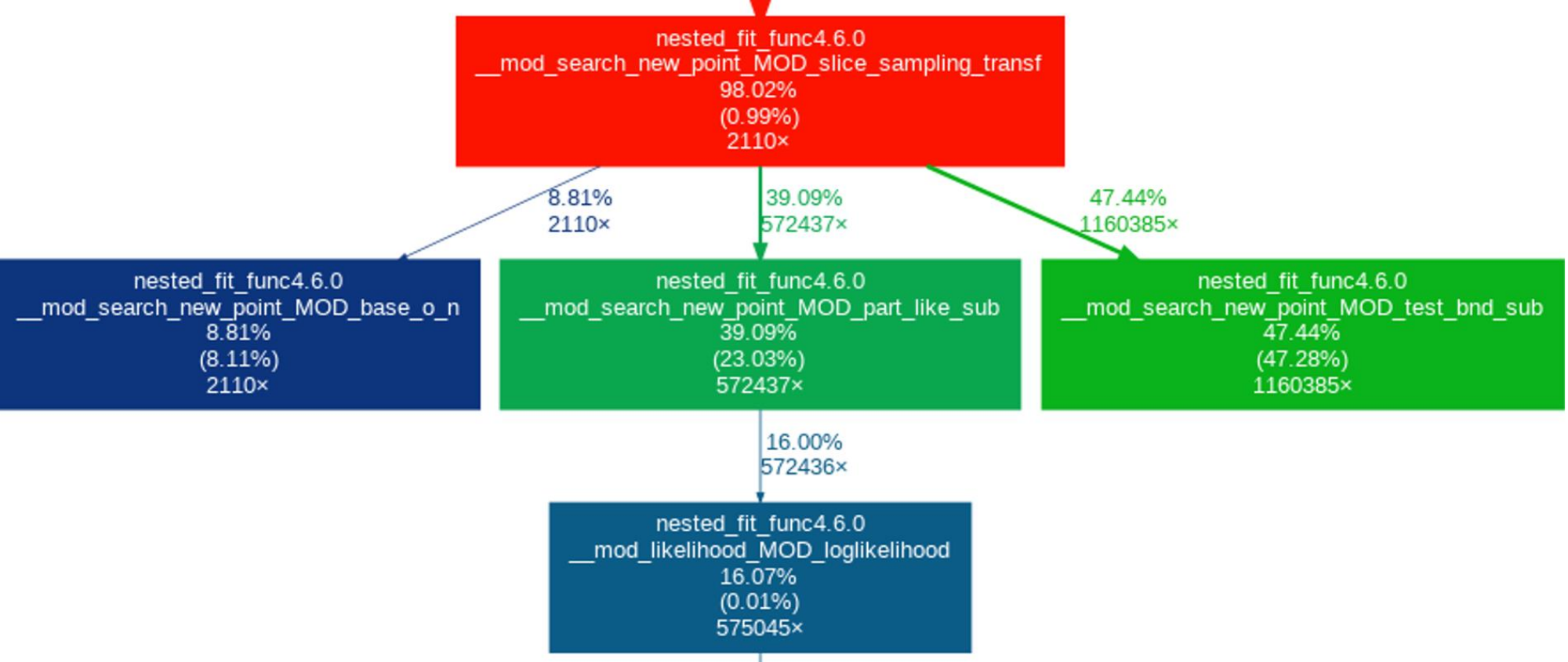}
\end{adjustwidth}
    \caption{Profiling for slice sampling transformed with the calculation of the covariance matrix every $0.05K$ iterations}
    \label{fig_profiling_sst_0.05K}    
\end{figure}
\begin{figure}
\begin{adjustwidth}{-\extralength}{0cm}
    \centering
    \includegraphics[scale=0.25,trim=0 100 0 0,clip]{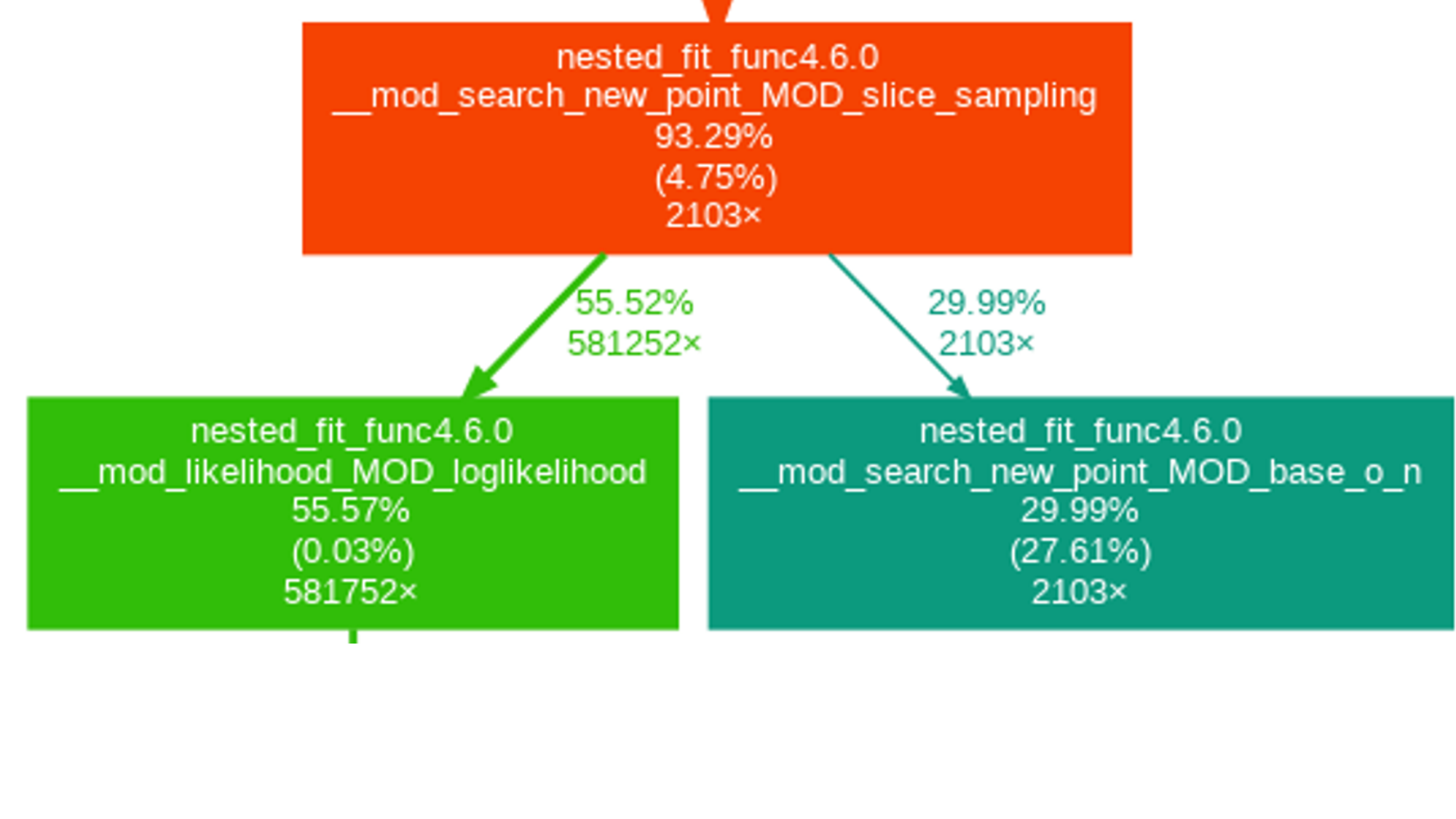}
\end{adjustwidth}
    \caption{Profiling for slice sampling real with the calculation of the covariance matrix every $0.05K$ iterations}
    \label{fig_profiling_ss_0.05K}    
\end{figure}

In this section, we look at the impact of the different developments made to \texttt{nested\_fit} on the computational resources needed. For that, we take the 36-atom cluster for its high number of parameters (108). First, we look at the result of profiling --- performed with \texttt{valgrind} \cite{noauthor_valgrind_nodate, nethercote_valgrind_2007} and \texttt{gprof2dot} \cite{noauthor_gprof2dot_nodate} ---, to see which parts of the algorithm are the most computationally expensive. We use $K=500$ and $n_\textrm{bases}=1$. We consider three cases: slice sampling transformed with the calculation of the covariance matrix at each iteration, slice sampling transformed with the calculation of the covariance matrix every $0.05K$ iterations and slice sampling real with the calculation of the covariance matrix every $0.05K$ iterations. The results of the profiling are presented in Figures \ref{fig_profiling_sst_full}, \ref{fig_profiling_sst_0.05K} and \ref{fig_profiling_ss_0.05K} respectively. In the first case, version 4.5.5 of \texttt{nested\_fit} is used and version 4.6.0 for the two other cases. %We used $K=500$ and $n_\textrm{bases}=1$ as, in this case, we want to compare the different implementation without necessarily having convergence.
The purpose of each function is explained in Table \ref{tab_profiling_expl}. Only the relevant boxes are represented. The last three lines in a box correspond to (1) the percentage of time spent in this function and its children function, (2) the percentage of time spent in this function alone and (3) the number of times the function was called. The arrows linking one box to another represent the first (parent) function calling the second (child) with the two numbers being the percentage of time the child function transfers to its parent and the number of times the parent function calls the child function. A box is only represented if at least 5\% of the time is spent in the function.

First, looking at Figures \ref{fig_profiling_sst_full} and \ref{fig_profiling_sst_0.05K}, we can see that, in the case of slice sampling transformed, over 50\% of the time is spent going back to the real space to calculate the likelihood and check the boundaries (i.e. in the part\_like\_sub and test\_bnd\_sub functions) which is not needed in slice sampling real. Thus, more time is spent changing space than calculating the function studied, leading to a computational overhead for slice sampling transformed. Furthermore, we can see that when the covariance function is computed at each iteration (Figure \ref{fig_profiling_sst_full}), this calculation takes around 20\% of the time while when it is only computed every $0.05K$ iterations (here every 25 iterations) (Figures \ref{fig_profiling_sst_0.05K} and \ref{fig_profiling_ss_0.05K}), the box does not appear, which means that less than 5\% of the time is spent on this particular task. Finally, we see that, for slice sampling transformed with the covariance matrix computed at each iteration (Figure \ref{fig_profiling_sst_full}), around 12\% of the time is spent calculating the function studied. This number goes up to around 55\% (4.4 times more) when using slice sampling real and computing the covariance function every $0.05K$ iterations (Figure \ref{fig_profiling_ss_0.05K}). In both cases, the number of times the energy function was called is roughly the same (around 580000 times). This shows that, by performing slice sampling in the real space rather than in the transformed space and not computing the covariance matrix at every iteration, we removed calculations that are not needed and that bring a significant computational overhead: the computational time is reduced by a factor 2.8, as can be deduced from Table \ref{tab_time_profiling_lj_36} which gives the computational time of all three cases for 20000 iterations.

For the impact of the parallelisation on the computation time, we also look at the time required to run \texttt{nested\_fit} with and without
parallelisation using slice sampling transformed with $K=1000$ and $n_\textrm{bases}=1$. The covariance matrix is computed every iteration. The parallelised case over 64 cores is around 21 times faster than the non-parallelised case (the non-parallelised case takes 85 minutes while the parallelised case takes 4 minutes). A possible explanation for this dividing factor being smaller than the number of cores used is that the time gained by simultaneously searching 64 new points may be partially counterbalanced by the time taken for the sequential adding of all the points, some of which may be rejected due to the constraint changing with each added point.

\begin{table}
    \caption{Purpose of the functions shown in Figures \ref{fig_profiling_sst_full} to \ref{fig_profiling_ss_0.05K}.}
    \label{tab_profiling_expl}
    \centering
    \begin{tabular}{|c|}
     \hline
     \textbf{\_mod\_search\_new\_point\_MOD\_slice\_sampling} \\
     Performs one iteration of nested sampling with slice sampling real \\
     \hline
     \textbf{\_mod\_search\_new\_point\_MOD\_slice\_sampling\_transf} \\
     Performs one iteration of nested sampling with slice sampling transformed\\
     \hline
     \textbf{\_mod\_likelihood\_MOD\_loglikelihood} \\
     Calls the function to calculate (here the Lennard-Jones potential) \\
     \hline
     \textbf{\_mod\_search\_new\_point\_MOD\_mat\_cov} \\
     Computes the covariance matrix \\
     \hline
     \textbf{\_mod\_search\_new\_point\_MOD\_base\_o\_n} \\
     Creates an orthonormal basis \\
     \hline
     \textbf{\_mod\_search\_new\_point\_MOD\_part\_like\_sub} \\
     Calculates the function for a point in the transformed space \\
     \hline
     \textbf{\_mod\_search\_new\_point\_MOD\_test\_bnd\_sub} \\
     Checks if a point in the transformed space is inside the sampling space \\
     \hline
    \end{tabular}
\end{table}
% \begin{table}
%     \caption{Purpose of the functions shown in Figures \ref{fig_profiling_sst_full} to \ref{fig_profiling_ss_0.05K}.}
%     \label{tab_profiling_expl}
%     \centering
% \begin{adjustwidth}{-\extralength}{0cm}
%      \begin{tabular}{c|c}
%      Name of the function & Purpose \\
%      \hline
%      \textbf{\_mod\_search\_new\_point\_MOD\_slice\_sampling} & Performs one iteration of nested sampling \\
%       & with slice sampling real \\
%      \hline
%      \textbf{\_mod\_search\_new\_point\_MOD\_slice\_sampling\_transf} & Performs one iteration of nested sampling \\
%       & with slice sampling transformed\\
%      \hline
%      \textbf{\_mod\_likelihood\_MOD\_loglikelihood} & Calls the function to calculate (here the Lennard-Jones potential) \\
%      \hline
%      \textbf{\_mod\_search\_new\_point\_MOD\_mat\_cov} & Computes the covariance matrix \\
%      \hline
%      \textbf{\_mod\_search\_new\_point\_MOD\_base\_o\_n} & Creates an orthonormal basis \\
%      \hline
%      \textbf{\_mod\_search\_new\_point\_MOD\_part\_like\_sub} & Calculates the function for a point in the transformed space \\
%      \hline
%      \textbf{\_mod\_search\_new\_point\_MOD\_test\_bnd\_sub} & Checks if a point in the transformed space is inside the sampling space \\
%      \hline
%     \end{tabular}   
% \end{adjustwidth}
% \end{table}

\begin{table}
    \caption{Time taken by all three cases to perform 20000 iterations without parallelisation.}
    \label{tab_time_profiling_lj_36}
    \centering
    \begin{tabular}{c|c|c|c}
      & Version 4.5.5 - Slice & Version 4.6.0 - Slice & Version 4.6.0 - Slice \\
      & sampling  transformed & sampling transformed & sampling real \\
     \hline
     Time (s) & 442.19 & 367.57 & 158.13
    \end{tabular}
\end{table}

\section{Conclusion}
In this work, we have applied the nested sampling algorithm to study classical Lennard-Jones clusters.

First, we have studied clusters of two sizes. As for the 7-atom cluster, we have seen that we were able to recover phase transitions by comparing our work with that in Ref. \cite{partay_efficient_2010}. We saw that we required more live points and more likelihood calls that in Ref. \cite{partay_efficient_2010}, likely due to the different choices of method to find a new point, which is an inconvenience for a more computationally expensive energy function. We then studied the 36-atom cluster which presents a solid-solid phase transition at low temperature that \texttt{nested\_fit} was able to recover provided enough live points are used.

We have then looked at the impact of different implementation on the computational resources: we have seen that using slice sampling transformed brought a consequent computational overhead through the need to transform back to the real space to compute the energy. Slice sampling real, which works directly in the real space, does not need to perform this task and therefore spends more than triple the time calculating the energy compared to slice sampling transformed, even though the function was called around the same number of times. We also saw that, using parallelisation on 64 cores, we gained a factor 21 in the computation time. All of those developments allowed us to study bigger and more complex clusters that require a higher number of live points. For example, one run for the 36-atom Lennard-Jones clusters with $K=70000$ took around 43 hours to run with the parallelisation.

In the future, we would like to study quantum Lennard-Jones clusters which are by about an order of magnitude more computationally expensive than the classical ones.

%%%%%%%%%%%%%%%%%%%%%%%%%%%%%%%%%%%%%%%%%%
\vspace{6pt} 

%%%%%%%%%%%%%%%%%%%%%%%%%%%%%%%%%%%%%%%%%%
%% optional
%\supplementary{The following supporting information can be downloaded at:  \linksupplementary{s1}, Figure S1: title; Table S1: title; Video S1: title.}

% Only for journal Methods and Protocols:
% If you wish to submit a video article, please do so with any other supplementary material.
% \supplementary{The following supporting information can be downloaded at: \linksupplementary{s1}, Figure S1: title; Table S1: title; Video S1: title. A supporting video article is available at doi: link.}

% Only for journal Hardware:
% If you wish to submit a video article, please do so with any other supplementary material.
% \supplementary{The following supporting information can be downloaded at: \linksupplementary{s1}, Figure S1: title; Table S1: title; Video S1: title.\vspace{6pt}\\
%\begin{tabularx}{\textwidth}{lll}
%\toprule
%\textbf{Name} & \textbf{Type} & \textbf{Description} \\
%\midrule
%S1 & Python script (.py) & Script of python source code used in XX \\
%S2 & Text (.txt) & Script of modelling code used to make Figure X \\
%S3 & Text (.txt) & Raw data from experiment X \\
%S4 & Video (.mp4) & Video demonstrating the hardware in use \\
%... & ... & ... \\
%\bottomrule
%\end{tabularx}
%}

%%%%%%%%%%%%%%%%%%%%%%%%%%%%%%%%%%%%%%%%%%
\authorcontributions{Conceptualisation, L.M., M.T. and F.F.; methodology, L.M., M.T. and F.F.; software, M.T., L.M. and C.G.; validation, L.M.; formal analysis, L.M.; investigation, L.M.; resources, F.F.; data curation, L.M. and M.T.; writing---original draft preparation, L.M.; writing---review and editing, F.F. and M.T..; visualisation, L.M.; supervision, M.T. and F.F.. All authors have read and agreed to the published version of the manuscript.}

\funding{This research received no external funding.}

%\informedconsent{Any research article describing a study involving humans should contain this statement. Please add ``Informed consent was obtained from all subjects involved in the study.'' OR ``Patient consent was waived due to REASON (please provide a detailed justification).'' OR ``Not applicable'' for studies not involving humans. You might also choose to exclude this statement if the study did not involve humans.

%Written informed consent for publication must be obtained from participating patients who can be identified (including by the patients themselves). Please state ``Written informed consent has been obtained from the patient(s) to publish this paper'' if applicable.}

\dataavailability{Raw data can be provided on demand.} 

% Only for journal Nursing Reports
%\publicinvolvement{Please describe how the public (patients, consumers, carers) were involved in the research. Consider reporting against the GRIPP2 (Guidance for Reporting Involvement of Patients and the Public) checklist. If the public were not involved in any aspect of the research add: ``No public involvement in any aspect of this research''.}

% Only for journal Nursing Reports
%\guidelinesstandards{Please add a statement indicating which reporting guideline was used when drafting the report. For example, ``This manuscript was drafted against the XXX (the full name of reporting guidelines and citation) for XXX (type of research) research''. A complete list of reporting guidelines can be accessed via the equator network: \url{https://www.equator-network.org/}.}

% Only for journal Nursing Reports
%\useofartificialintelligence{Please describe in detail any and all uses of artificial intelligence (AI) or AI-assisted tools used in the preparation of the manuscript. This may include, but is not limited to, language translation, language editing and grammar, or generating text. Alternatively, please state that “AI or AI-assisted tools were not used in drafting any aspect of this manuscript”.}

\acknowledgments{Work realised with the support of the Sorbonne Center for Artificial Intelligence 579
- Sorbonne University - IDEX SUPER 11-IDEX-0004. The authors thank Philipe Depondt, Simon Huppert and Julien Salomon for helpful discussions.}

\conflictsofinterest{The authors declare no conflicts of interest.} 

%%%%%%%%%%%%%%%%%%%%%%%%%%%%%%%%%%%%%%%%%%
%% Optional

%% Only for journal Encyclopedia
%\entrylink{The Link to this entry published on the encyclopedia platform.}

%%%%%%%%%%%%%%%%%%%%%%%%%%%%%%%%%%%%%%%%%%
\begin{adjustwidth}{-\extralength}{0cm}
%\printendnotes[custom] % Un-comment to print a list of endnotes

\reftitle{References}

% Please provide either the correct journal abbreviation (e.g. according to the “List of Title Word Abbreviations” http://www.issn.org/services/online-services/access-to-the-ltwa/) or the full name of the journal.
% Citations and References in Supplementary files are permitted provided that they also appear in the reference list here. 

%=====================================
% References, variant A: external bibliography
%=====================================
\bibliography{Bibliography.bib}

%=====================================
% References, variant B: internal bibliography
%=====================================

% If authors have biography, please use the format below
%\section*{Short Biography of Authors}
%\bio
%{\raisebox{-0.35cm}{\includegraphics[width=3.5cm,height=5.3cm,clip,keepaspectratio]{Definitions/author1.pdf}}}
%{\textbf{Firstname Lastname} Biography of first author}
%
%\bio
%{\raisebox{-0.35cm}{\includegraphics[width=3.5cm,height=5.3cm,clip,keepaspectratio]{Definitions/author2.jpg}}}
%{\textbf{Firstname Lastname} Biography of second author}

% For the MDPI journals use author-date citation, please follow the formatting guidelines on http://www.mdpi.com/authors/references
% To cite two works by the same author: \citeauthor{ref-journal-1a} (\citeyear{ref-journal-1a}, \citeyear{ref-journal-1b}). This produces: Whittaker (1967, 1975)
% To cite two works by the same author with specific pages: \citeauthor{ref-journal-3a} (\citeyear{ref-journal-3a}, p. 328; \citeyear{ref-journal-3b}, p.475). This produces: Wong (1999, p. 328; 2000, p. 475)

%%%%%%%%%%%%%%%%%%%%%%%%%%%%%%%%%%%%%%%%%%
%% for journal Sci
%\reviewreports{\\
%Reviewer 1 comments and authors’ response\\
%Reviewer 2 comments and authors’ response\\
%Reviewer 3 comments and authors’ response
%}
%%%%%%%%%%%%%%%%%%%%%%%%%%%%%%%%%%%%%%%%%%
\PublishersNote{}
\end{adjustwidth}
\end{document}